\documentclass{aa}  
\usepackage[varg]{txfonts}
\usepackage[colorlinks=true,linkcolor=blue,citecolor=blue]{hyperref}
\usepackage{graphicx}
\usepackage{tikz}
\usepackage{color}

\begin{document}

   \title{Increasing the raw contrast of VLT/SPHERE with the dark hole technique.\\I. Simulations and validation on the internal source}

   \author{A. Potier\inst{1}
          \and
          R. Galicher\inst{1}
          \and
          P. Baudoz\inst{1}
          \and
          E. Huby\inst{1}
          \and
          J. Milli\inst{2,}\inst{3}
          \and
          Z. Wahhaj\inst{3}
          \and
          A. Boccaletti\inst{1}
          \and
          A. Vigan\inst{4}
          \and
          M. N'Diaye\inst{5}
          \and
          J.-F. Sauvage\inst{4,} \inst{6}
          }

   \institute{\inst{1} LESIA, Observatoire de Paris, Université PSL, CNRS, Sorbonne Université, Université de Paris, 5 place Jules Janssen, 92195 Meudon, France\\
   \inst{2} Univ. Grenoble Alpes, CNRS, IPAG, F-38000 Grenoble, France\\
   \inst{3} European Southern Observatory, Alonso de Cordova 3107, Vitacura, Santiago, Chile\\
   \inst{4} Aix Marseille Univ, CNRS, CNES, LAM, Marseille, France \\
   \inst{5} Université Côte d'Azur, Observatoire de la Côte d'Azur, CNRS, Laboratoire Lagrange, France \\
   \inst{6} ONERA, The French Aerospace Lab, BP72, 29 Avenue de la Division Leclerc, 92322 Châtillon Cedex, France \\
              \email{axel.potier@obspm.fr}
             }
   \date{Received 24 March 2020 ; accepted 27 April 2020}

 
  \abstract
   {Since 1995 and the first discovery of an exoplanet orbiting a main-sequence star, 4000 exoplanets have been discovered using several techniques. However, only a few of these exoplanets were detected through direct imaging. Indeed, the imaging of circumstellar environments requires high-contrast imaging facilities and accurate control of wavefront aberrations. Ground-based planet imagers such as VLT/SPHERE or Gemini/GPI have already demonstrated great performance. However, their limit of detection is hampered by suboptimal correction of aberrations unseen by adaptive optics (AO).}
   {Instead of focusing on the phase minimization of the pupil plane as in standard AO, we aim to directly minimize the stellar residual light in the SPHERE science camera behind the coronagraph to improve the contrast as close as possible to the inner working angle.}
   {We propose a dark hole (DH) strategy optimized for SPHERE. We used a numerical simulation to predict the global improvement of such a strategy on the overall performance of the instrument for different AO capabilities and particularly in the context of a SPHERE upgrade. Then, we tested our algorithm on the internal source with the AO in closed loop.}
   {We demonstrate that our DH strategy can correct for aberrations of phase and amplitude. Moreover, this approach has the ability to strongly reduce the diffraction pattern induced by the telescope pupil and the coronagraph, unlike methods operating at the pupil plane. Our strategy enables us to reach a contrast of 5e-7 at 150 mas from the optical axis in a few minutes using the SPHERE internal source. This experiment establishes the grounds for implementing the algorithm on sky in the near future. 
   }
   {}

   \titlerunning{Increasing the raw contrast of VLT/SPHERE with the dark hole technique.}
   \maketitle
\section{Introduction}
High-contrast imaging (HCI) is a powerful technique to detect substellar companions down to the planetary mass regime and to perform the characterization of their atmospheres with spectroscopy. Ground-based instruments such as VLT/SPHERE \citep{Beuzit2019} and Gemini/GPI \citep{Macintosh2015} have led to tens of scattered-light images of circumstellar disks \citep[e.g., ][]{Hung2015, Kalas2015, Perrot2016,Sissa2018}, new discoveries of young and massive exoplanets \citep{Macintosh2015Science, Chauvin2017, Keppler2018}, and have allowed for the constraint of physical properties of already known objects \citep[e.g., ][]{Boccaletti2018,Bhowmik2019}. However, imaging exoplanets from the ground faces several challenges because of the apparent proximity between these objects and their much brighter host stars.
First, the instrument requires high angular resolution to resolve planetary system scales -- a few astronomical unit (AU) to a few tens of AU -- for nearby stars. Second, a current coronagraph with a small inner working angle (IWA) associated with an extreme adaptive optics (AO) system is mandatory to reject the starlight so that its faint environment (exoplanets, disks) can be imaged.\\

Coronagraphs provide their best performance in the absence of aberrations. Unfortunately, even the most powerful AO systems do not compensate for all aberrations. For example, these systems do not correctly handle non-common-path aberrations \citep[NCPAs;][]{Fusco2006}. These NCPAs result from the difference of the optical path after the beam splitter between the science path and the wavefront sensor path. The NCPA level is typically $\sim$50 nm root mean square (rms) over the pupil in SPHERE \citep{Vigan2019} and GPI instruments. These NCPAs cause stellar speckles that mimic exoplanet images in the coronagraphic science image. Several post-processing techniques, such as spectral differential imaging \citep[SDI;][]{Racine1999,Marois2000}, polarimetric differential imaging \citep[PDI;][]{Kuhn2001} or angular differential imaging \citep[ADI;][]{Marois2006}, have been designed to calibrate part of these speckles and to improve the contrast typically by a factor of about 10. To go one step beyond the abilities of HCI and to detect fainter exoplanets closer to the star, the NCPAs should be corrected beforehand during the target acquisition. Since NCPAs are quasi-static, meaning they are slowly changing in time owing to thermal and mechanical variations along with turbulence inside the instrument, they have to be compensated regularly during the night at a frequency that depends on their lifetime. An efficient strategy to compensate for the NCPAs is to directly estimate the aberrations from the science detector using a focal plane wavefront sensor (FPWFS). A few sensors have already been implemented and validated on optical test beds fed by an artificial residual turbulence \citep{Singh2019, PotierAO4ELT2019, Herscovici2019}. Some of these sensors were tested in the past to calibrate the quasi-static aberrations on calibration sources: the self-coherent camera \citep[SCC;][]{Galicher2019} and the electric field conjugation \citep[EFC;][]{Matthews2017} were tested at the Palomar Observatory, the coronagraphic phase diversity \citep[COFFEE;][]{Paul2014} and the Zernike sensor for extremely low-level differential aberration \citep[ZELDA;][]{NDiaye2016} were studied on SPHERE, while the speckle nulling technique was implemented at Palomar and Keck \citep{Bottom2016}. However, only the speckle nulling technique with SCEXAO/Subaru \mbox{\citep{Martinache2014}} and ZELDA with SPHERE \citep{Vigan2019} were tested directly using on-sky measurements. These methods have shown moderate improvements in terms of raw contrast because of various limitations. 

In this paper, we propose to estimate the performance of dark hole (DH) techniques, which focus on minimizing the stellar intensity in a chosen region of the science detector \citep{Malbet1995}, applied on a current HCI instrument such as SPHERE. In Sect. \ref{sec:SPHERE}, we describe the SPHERE instrument and the different sources of contrast limitations. Then, we present in Sect. \ref{sec:WCwAPLC} our DH control strategy using a pair-wise (PW) wavefront sensor and an EFC controller. We numerically simulate the performance we expect to reach with these techniques under several conditions.  Finally, we depict the strategy used to apply PW+EFC on SPHERE and we demonstrate a full correction of the quasi-static aberrations with the SPHERE internal calibration unit in Sect. \ref{sec:WCinternalpupil}.

\section{Current SPHERE instrument setup and limitations}
\label{sec:SPHERE}
Our strategy of FPWFS and correction relies on a good knowledge of the instrument (see Sect. \ref{sec:WCwAPLC}). Thus, in this section we describe the hypotheses adopted for our model of the instrument. The SPHERE HCI is described in detail in \cite{Beuzit2019}. In the following, we only consider the subsystems common path and infrastructure (CPI) and the Infrared Dual Imager and Spectrograph (IRDIS).
        \subsection{Model of SPHERE}
        \label{subsec:SPHEREModel}
        
                        \subsubsection{Adaptive optics loop}
                \label{subsubsec:AOloop}
\begin{figure}
\centering
   \includegraphics[width=9cm]{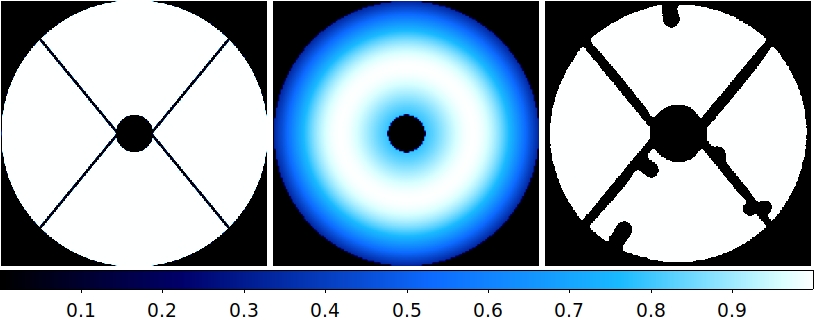}
   \caption[example] 
   { \label{fig:Model} Transmission maps of the entrance VLT pupil (left), the apodiser (center), and the Lyot stop (right) used in the APO1-ALC2 configuration.}
\end{figure}

The light coming from the target star and its environment reaches the VLT pupil that is represented in the left image of Fig.~\ref{fig:Model} with its central obstruction and spiders. Inside the instrument, the visible light is separated from the infrared light to feed a high order extreme AO \citep[SAXO;][]{Fusco2006}. The wavefront sensing of the AO is performed with a 1240 subpupil Shack-Hartman (SH) and an EMCCD detector with a low read-out noise (0.1 $e^{-}$ per pixel). The positions of each spot of the SH are measured thanks to a weighted center of gravity algorithm and these are used to estimate the local wavefront. In order to remove aliasing effects on the reconstructed wavefront, a filtering squared pinhole can be adjusted as a function of the atmospheric conditions. In the tests described in this paper (Sect.~\ref{sec:WCinternalpupil}), the filter is set to MEDIUM, which corresponds to a size of $\sim$1.3~arcseconds.\\
The wavefront perturbation is corrected thanks to a fast image tip-tilt mirror (ITTM) whose bandwidth is 800Hz and a high order deformable mirror (HODM) made of 41x41 actuators, and working at 1380Hz. 1377 actuators are located in the pupil but at least six of these are faulty. 

                \subsubsection{Infrared light path}
                \label{subsubsec:lightpath}

    The infrared light goes through the instrument to a stellar coronagraph. We consider only one configuration of the apodized pupil lyot coronagraph \citep[APLC;][]{Carbillet2011} working in the H band because it is the most frequently used mode of SPHERE during the large survey \citep{Chauvin2017SF2A}. The coronagraph is composed of a pupil apodizer followed by a focal plane mask (FPM) and a Lyot stop. The apodizer, named APO1, minimizes the starlight diffracted in the final IRDIS image when associated with the FPM and the Lyot stop. The transmission of APO1 is represented in Fig.~\ref{fig:Model} (center). The FPM is an opaque disk of 185 mas diameter (named ALC2). Finally, the Lyot stop is a binary mask that is set in a pupil plane and represented in Fig. \ref{fig:Model} (right). The shape of the Lyot stop was designed to undersize the telescope pupil to 96\% at the outer edge, while the central obscuration and telescope spiders are made larger than in the full pupil (respectively 20\% and 2.5\%). On top of it, six patches that are about 5\% of the radius of the pupil are intended to block the light diffracted by the six defective actuators in the deformable mirror (see Sect. \ref{subsubsec:AOloop}). The overall transmission of the APLC for an off-axis source is 58\% in this particular configuration (APO1-ALC2).\\

        \subsection{Limitations in contrast}
        \label{subsec:SPHERELimitations}
The current SPHERE performance is decribed in \cite{Cantalloube2019}. The performance is mainly limited by the capability of the AO system to correct for the wind-driven halo \citep{Cantalloube2018}, low-order residuals, low-wind effect \citep{Sauvage2015, Milli2018}, NCPAs \citep[][]{NDiaye2016}. The first three limitations require improving the AO system or  observing during excellent weather conditions. The current instrument can, to some extent,  compensate for the NCPAs.

\begin{figure}
\centering

\begin{tikzpicture}
        \node (tiger) [anchor=south west, inner sep=0pt] {\includegraphics[width=8.3cm]{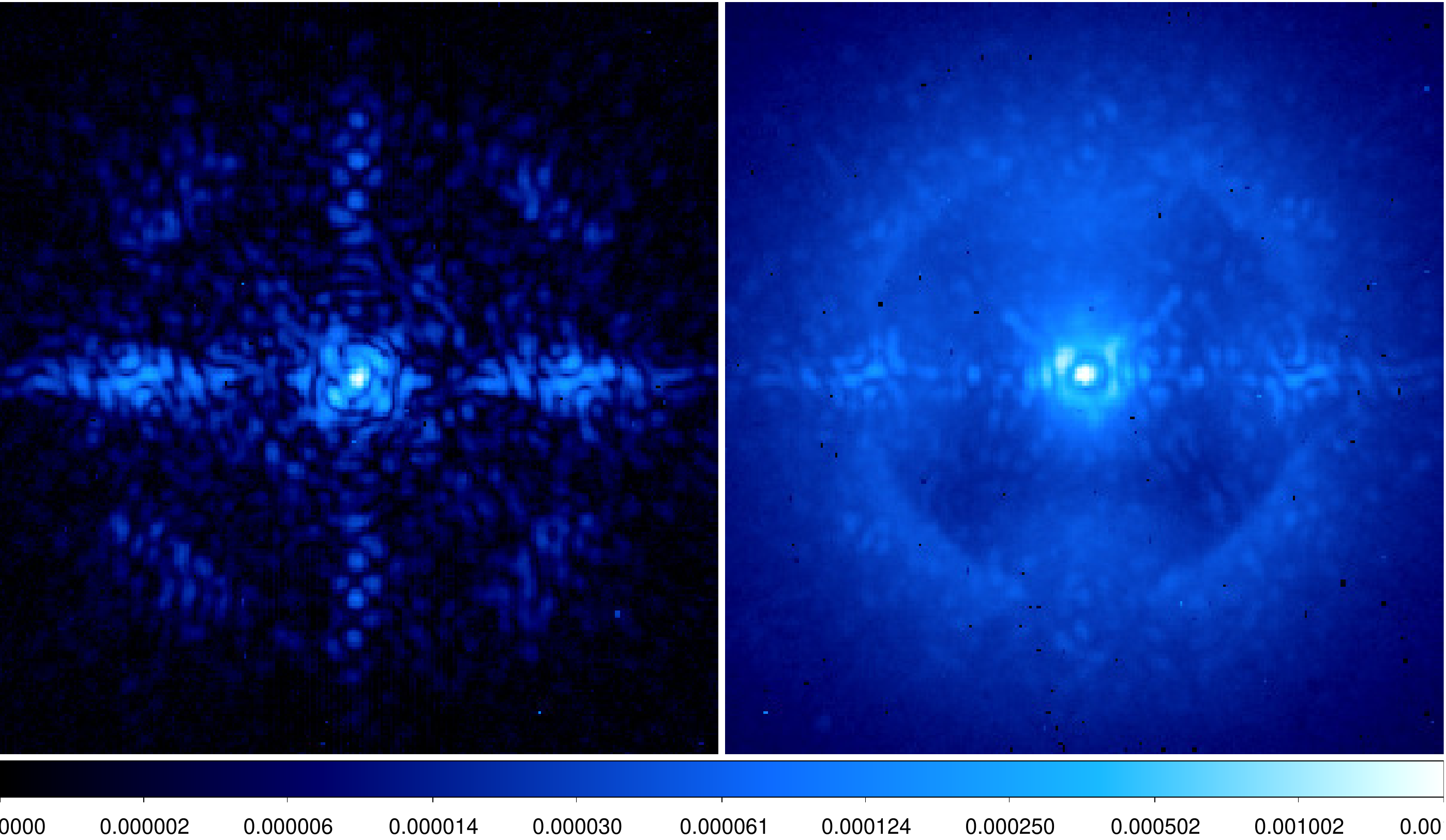}};
        \begin{scope}[x={(tiger.south east)},y={(tiger.north west)}]
        \fill[white] (4.5cm,0cm) rectangle (8.4cm,0.5cm);
    \end{scope}
    \begin{scope}[x={(tiger.south west)},y={(tiger.north east)}]
        \fill[white] (0cm,0cm) rectangle (5cm,0.5cm);
    \end{scope}
      \end{tikzpicture}
   \caption[example] 
   { \label{fig:ImageInternalPupAndOnSky} 
   IRDIS images in the H3 bandwidth ($\lambda=1667nm$, $\Delta\lambda=54nm$) recorded with the internal calibration unit (left) and on sky on a bright star under 0.65$''$ seeing conditions (right).}
\end{figure}

We represent in Fig. \ref{fig:ImageInternalPupAndOnSky} two raw IRDIS images taken with the internal source calibration and with a bright source on sky under good seeing conditions (about 0.65 arcsec). Both images were recorded while the AO was in closed-loop with no compensation of the NCPA. On the left image with no turbulence, residual speckles are present inside the corrected region. The size of each speckle is typically one resolution element and limits the raw contrast to $10^{-4}$ at 200 mas from the optical axis. These speckles define an ultimate floor for the detection of point sources in the raw image when the seeing conditions are good (see in Fig. \ref{fig:ImageInternalPupAndOnSky}, right). The ZELDA wavefront sensor has proved to be efficient to compensate for most of the NCPAs \citep{Vigan2018}. However, the improvement on the contrast is limited by the residual diffracted light of the coronagraph at the image center, amplitude aberrations, and some phase residuals \citep{Vigan2019}.\\

Currently, these speckles can partially be filtered out by post-processing techniques such as ADI. However, differential imaging requires the stability of the aberrations within an observing sequence. Thus, any quasi-static speckles varying at a fraction of the observing time are not optimally removed. As for SDI, the performance is limited by the chromaticity of aberrations. By design, the ADI efficiency degrades rapidly at close angular separations. For these reasons, the gain in contrast of both ADI and SDI is at maximum 10, resulting in detection limit in the range of 10$^{-5}$ to 10$^{-6}$ inside the control radius of SAXO \citep{Beuzit2019}. 
In addition, both techniques cause self-subtraction, which biases the photometry and astrometry of point sources \citep{Galicher2011}; this bias is even worse for extended objects such as disks \citep{Milli2012}. For all these reasons, it is necessary to minimize the speckle level in the science images during the observations, and not only a posteriori as done nowadays.

        \subsection{Description of IRDIS electric field}
        \label{subsec:IRDISElectricField}
This section presents the mathematical model that explains the coronagraphic intensity as a function of the configuration and aberrations of the instrument. We call $C$ the linear operator, which transforms the complex electric field $E$ in the entrance pupil to the electric field in the IRDIS detector plane. We can assume Fourier propagation and write $C$ as
\begin{equation}
\label{eq:propmodel}
\begin{aligned}
C(E)= &\mathcal{F}\left[\mathcal{F}^{-1} \left[\text{FPM}\times \mathcal{F}[E]\right]\times \text{L}\right]\\
=& \left[\text{FPM}\times\mathcal{F}[E]\right]*\mathcal{F}[\text{L}]
\end{aligned}
,\end{equation}  
where $*$ is the convolution symbol, $\mathcal{F}$ and $\mathcal{F}^{-1}$ denotes the direct and inverse Fourier transform operators respectively, FPM represents the occulting FPM, and L is the Lyot stop (Fig.~\ref{fig:Model}, right).

We consider only static aberrations. The electric field in the entrance pupil is
\begin{equation}
\label{eq:totalelectricfield}
E=Ae^{\alpha+i\beta}e^{i\phi},
\end{equation}
where $\alpha$ and $\beta$ are the log-amplitude and phase aberrations in the instrument pupil plane. The quantity $\phi$ is the phase induced by the HODM+TT and $A$ is the unaberrated electric field in the pupil plane such as
\begin{equation}
\label{eq:EntrancePupil}
A =\left\{
    \begin{array}{ll}
        \text{P}\times \text{APO} & \mbox{ when observing on sky}\\
        \text{APO} & \mbox{ when using the internal source,}
        \end{array}
        \right.
\end{equation}
where P represents the geometry of the VLT pupil with the central obscuration and the spiders and APO denotes the apodizer transmission. If we assume small upstream aberrations and small deformations of the DM, we can write the Taylor expansion of Eq.~\ref{eq:totalelectricfield} to describe the electric field $E_{IRDIS}$ in the IRDIS science detector plane as follows:
\begin{equation}
\label{eq:totalelectricfield2}
\begin{aligned}
E_{IRDIS}&=C[E]\\
&=C[A]+C\left[A(\alpha+i\beta)\right]+iC\left[A\phi\right]\\
&= E_{D}+E_{S}+E_{DM}
\end{aligned}
,\end{equation}
where $E_D$ is the diffracted field in the IRDIS detector plane in the absence of any aberration. A perfect coronagraph would null $E_D$, but the APLC used on SPHERE leaves a residual diffracted light that could limit the contrast especially at short angular separations. The quantities $E_{S}$ and $E_{DM}$ are the focal plane electric field corresponding to the diffraction of the small upstream initial aberrations (both in phase $\beta$ and amplitude $\alpha$) and the small deformation of the DM (creating a phase $\phi$) through the coronagraph, respectively. In the presence of static aberrations, part of the stellar light goes through the system and reaches the science detector inducing stellar speckles. The quantity $E_S$ is linked to the stellar speckles and $E_{DM}$ is used to minimize $E_S$.
Wavefront sensors only trying to estimate NCP phase aberrations can only minimize $\beta$. It is thus crucial to measure the complete electric field $E_{IRDIS}$ and then apply a control strategy to minimize $E_{IRDIS}$ by compensating $E_D$ and $E_S$ with $E_{DM}$ using the HODM. This is the goal of DH techniques.

\section{Wavefront control to generate a dark hole with SPHERE}
\label{sec:WCwAPLC}
Creating a DH at the focal plane involves two independent steps. First, the electric field is estimated with a FPWFS. Then, the HODM is used to minimize the intensity inside the DH. In the following we describe how these two steps are implemented using PW probing for measuring the electric field and then the EFC for computing the commands to apply on the HODM.
        \subsection{Pair-wise probing}
        \label{subsec:PW}
The incident electric field on the IRDIS camera cannot be estimated well by the WFS of the AO system because it does not measure amplitude aberrations and NCPAs, which includes defects of the coronagraph. That is why we need to estimate the aberrations directly from the science focal plane. However the science detector only records the square modulus of the electric field. Temporally modulating the speckle field is a solution to measuring the complex focal plane electric field. Such a technique has already been tested on SPHERE. For instance, COFFEE was performed by \cite{Paul2014} but it was only estimating phase aberrations upstream and downstream the coronagraph FPM. Another algorithm called PW probing was also tested. This algorithm demonstrated an accurate estimation of the electric field in a small area of the focal plane \citep{Ruffio2014,Fusco2015}. We propose a PW solution to increase the area in which the electric field is estimated well.
First described by \cite{GiveOn2007SPIE}, PW requires the introduction of known aberrations called probes thanks to the HODM. Still assuming small upstream aberrations and small probes $\psi_m$ introduced by the deformable mirror, the intensity on IRDIS detector can be written as
\begin{equation}
I_m=|E_{D}+E_{S}+iC[A\psi_m]|^2.
\end{equation}
Recording a pair of images $I_m^+$ and $I_m^-$ for which the probes are $+\psi_m$ and $-\psi_m$, respectively, we can compute the difference between the two images as follows:
\begin{equation}
\label{eq:DifferenceResult}
I_m^+-I_m^-=4(\Re(E_{S}+E_{D})\Re(iC[A\psi_m])+\Im(E_{S}+E_{D})\Im(iC[A\psi_m])),\end{equation}
where $\Re$ and $\Im$ represent the real and imaginary parts of a complex field. Noticing that Eq.~\ref{eq:DifferenceResult} can be written in the matrix notation
\begin{eqnarray}
\label{eq:DifferenceResult2} 
I_m^+-I_m^-=4\begin{bmatrix}
\Re (iC[A\psi_m]) & \Im (iC[A\psi_m])
\end{bmatrix} \begin{bmatrix}
 \Re (E_{S}+E_{D}) \\
 \Im (E_{S}+E_{D})
\end{bmatrix},
\end{eqnarray}
we can generalize this principle for $k$ different probes to break the degeneracy between the real and imaginary part of $E_S+E_D$. Then, we write, for each pixel with coordinates $(x,y)$ in the science image,
\begin{eqnarray}
\label{eq:estimation} 
D_{(x,y)}=4M_{(x,y)}F_{(x,y)} .
\end{eqnarray}
The difference matrix $D$ represents the subtraction of the two images acquired for each pair of probes as follows:
\begin{eqnarray}
\label{eq:DifferenceMatrix}
D=\begin{bmatrix}
 I_1^+-I_1^- \\
. \\
. \\
. \\
I_k^+-I_k^-
\end{bmatrix}.
\end{eqnarray}
The model matrix $M$ is composed of the numerically simulated electric fields created on IRDIS detector by applying each probe on the HODM as follows:
\begin{eqnarray} 
\label{eq:ModelMatrix}
M=
\begin{bmatrix}
\Re (iC[A\psi_1]) & \Im (iC[A\psi_1]) \\
. & . \\
. & . \\
. & . \\
\Re (iC[A\psi_k]) & \Im (iC[A\psi_k]) 
\end{bmatrix}.
\end{eqnarray}
The propagation of the aberration through the instrument and the APLC is simulated according to \cite{Soummer2007}.  This accounts for the positions and influence functions of all the actuators with respect to the entrance pupil. Finally, the electric field matrix $F$ is composed of the electric field, created by the upstream aberrations and the diffraction pattern of the instrument, as follows:
\begin{eqnarray}
F=
\begin{bmatrix}
 \Re (E_{S}+E_{D}) \\
 \Im (E_{S}+E_{D})
\end{bmatrix} .
\end{eqnarray}
We aim to retrieve $E_{S}+E_{D}$ by inverting Eq.~\ref{eq:estimation} for each desired pixel $(x,y)$ in the field of view. However, we have to ensure the matrix $M$ is invertible, which means its determinant is non-zero. Therefore, PW needs at least two probes $\psi_m$ and $\psi_n$ so that they create a different electric field at the particular location $(x,y)$, that is,
\begin{eqnarray}
\label{eq:EnoughDiversity}
\Re(iC[A\psi_m])\Im(iC[A\psi_n])-\Re(iC[A\psi_n])\Im(iC[A\psi_m])\neq 0 .
\end{eqnarray}
We can then inverse Eq.~\ref{eq:estimation} for all the pixels $(x,y),$ where Eq.~\ref{eq:EnoughDiversity} is verified  to estimate the real and imaginary part of the electric field $E_S+E_D$, that is,
\begin{eqnarray}
\label{eq:estimation2}
\tilde{F}_{(x,y)}=\frac{1}{4}M^{\dagger}_{(x,y)}D_{(x,y)} ,
\end{eqnarray}
where $\tilde{F}$ represents the estimation of $F$ and $\dagger$ is the pseudo inverse, calculated with the singular value decomposition (SVD) method, as the matrix is not squared when using more than two probes.\\

\begin{figure}
\centering
\begin{tikzpicture}
        \node (tiger) [anchor=south west, inner sep=0pt] {\includegraphics[width=8.2cm]{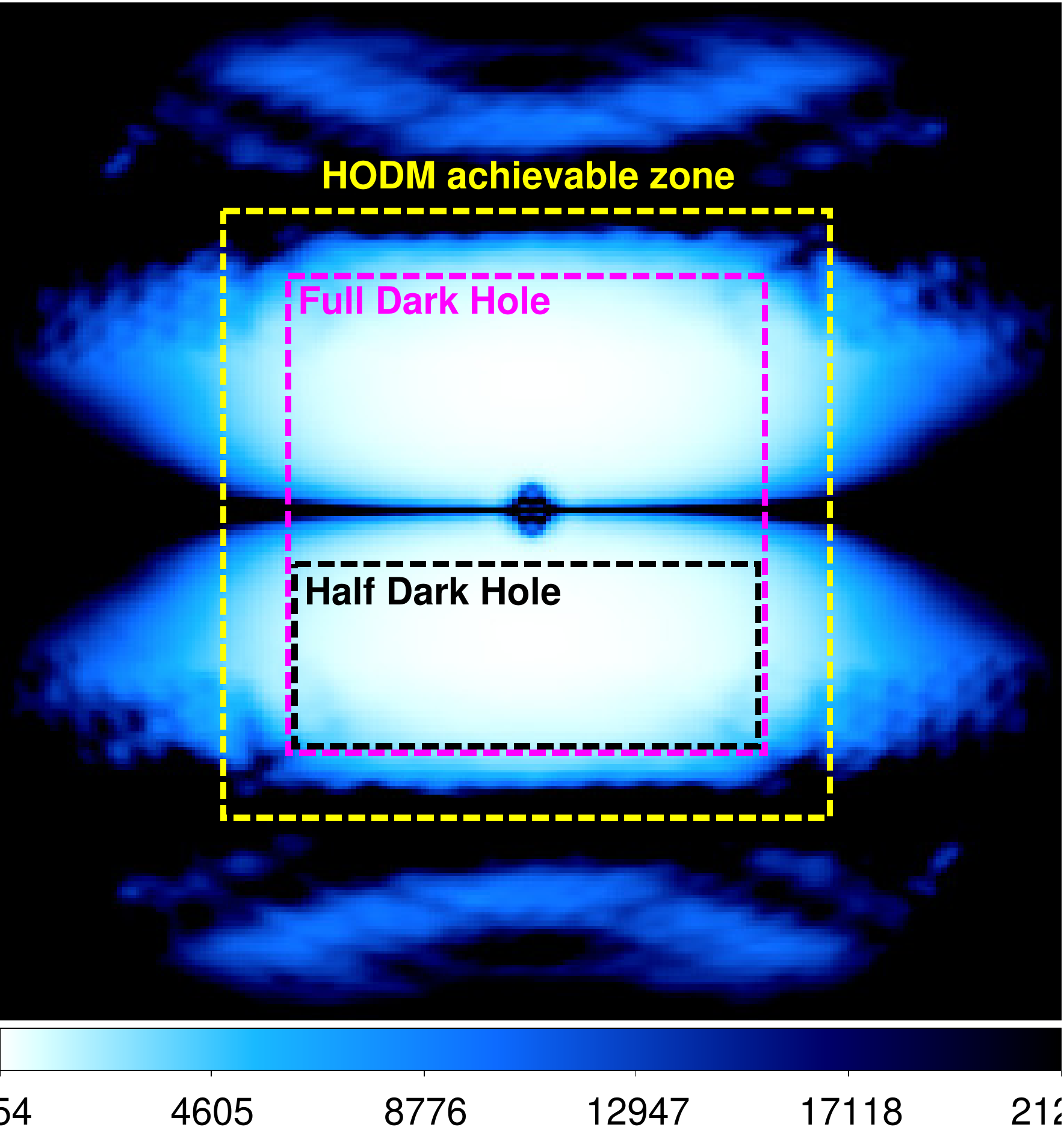}};
        \begin{scope}[x={(tiger.south east)},y={(tiger.north west)}]
        \fill[white] (7.8cm,0cm) rectangle (8.4cm,0.40cm);
    \end{scope}
    \begin{scope}[x={(tiger.south west)},y={(tiger.north east)}]
        \fill[white] (0cm,0cm) rectangle (0.5cm,0.40cm);
    \end{scope}
      \end{tikzpicture}
   \caption[example] 
   { \label{fig:InverseSVD} 
   Map of the inverse of the minimum eigenvalue for the matrix $M$ around the optical axis. The quantity $M$ is computed by simulating the effect of pushing two neighbor actuators aligned in the vertical direction. The HODM achievable zone, FDH, and HDH regions are also highlighted in yellow, pink, and black dashed lines, respectively.}
\end{figure}

The choice of the probe shapes and locations is important when implementing PW. Indeed, the foreseen strategy is to point a target star, use PW and EFC to minimize the speckle intensity and eventually record images for astrophysical purpose. To minimize the time spent for stellar speckle minimization, we aim to use the smallest number of probes while making sure Eq.~\ref{eq:EnoughDiversity} still holds for all the pixels in the DH where we seek to minimize the speckle intensity. We  demonstrate in \cite{Potier2020} that pushing and pulling two neighbor actuators allows us to estimate properly the electric field in almost the entire field of view. We therefore decided to use two actuator pokes as probes, aligned in the vertical direction and located on the HODM just below the Lyot stop obstruction shadow. To determine the accuracy of PW when using these two bumps, we show in Fig.~\ref{fig:InverseSVD} the inverse of the minimum eigenvalue of $M$ for each IRDIS pixel. As seen in \cite{Potier2020}, this map displays the regions in which $M$ is invertible. In Fig.~\ref{fig:InverseSVD}, the whiter the pixel, the more accurate the estimation of the electric field. This demonstrates that the two selected probes allow for a good estimation of the electric field in the entire HODM influence area (set by the number of actuators), except for a horizontal bar at the image center and on the edges of the HODM influence zone. We chose the zones in which we minimize the speckle field: a full DH (FDH) that is $31\lambda/D\times31\lambda/D$ and a half DH (HDH) that is $13\lambda/D\times31\lambda/D$, whose edge is located at 2.85$\lambda/D$ from the optical axis. In these regions, the inverses of the $M$ singular values do not seem to diverge -- except from the central horizontal bar in the FDH. If an observer wants to remove the horizontal bar to minimize the speckle field in a HDH in the right (or left) part of the coronagraphic image, the probes would also be two pokes of neighbor actuators, but aligned in the horizontal direction and located on the right or left part of the central obstruction.
        \subsection{Electric field conjugation}
        \label{subsec:EFC}
In this section, we present how to suppress speckles once the focal plane electric field is estimated by PW. In former tests for the compensation of NCPAs on SPHERE with COFFEE \citep{Paul2014} and ZELDA \citep{Vigan2018, Vigan2019}, the focal plane images were used to estimate only the phase aberrations reconstructed in the pupil plane. These aberrations were corrected by applying the opposite estimated optical path difference (OPD) on the HODM to flatten the wavefront. This strategy does not allow us to correct for the amplitude aberrations $e^\alpha$, which is no longer negligible at these levels of contrast \citep{Vigan2016}, and also cannot suppress the intensity of the diffracted residual diffraction pattern of the coronagraph $|E_D|^2$.\\

Since PW estimates the electric field in the IRDIS detector plane, we used a strategy proposed more than two decades ago for space-based applications, which consists in generating a DH in a chosen region of the science detector \citep{Malbet1995}. Using a single HODM in a pupil plane, we can either choose to minimize the speckle intensity induced by phase aberrations ($\beta$) in the full DM influence zone (full DH) or the intensity of all speckles induced by phase and amplitude aberrations, as well as the residual diffraction pattern $|E_D|^2$ in a half DH (HDH).\\

As formulated in \cite{Borde2006} and \cite{GiveOn2007}, the EFC is intended to attenuate the speckle field intensity by minimizing the following metric:
\begin{equation}
\label{eq:CostFunction}
d^2_{EFC}=||E_{DM}+(E_{S}+E_{D})||^2
\end{equation}
in all the pixels at once inside the defined DH. This least mean squares criteria can be minimized in the DH region by applying a phase $\phi$ on the HODM. We assume that the electric field created by the HODM in the focal plane $E_{DM}$ is a linear combination of the individual actuators voltages as follows:
\begin{equation}
\label{eq:LinEFC}
E_{DM}=G\bar{a} , 
\end{equation}
where $\bar{a}$ is the vector composed of the 1377 voltages of the HODM and $G$ is the linear transformation matrix between the voltage parameter and the focal plane electric field. The quantity $G$ is computed numerically by simulating the electric field created in the chosen DH region when adding a voltage unit to each of the actuators in the pupil, except for the six faulty actuators.\\
Eq.~\ref{eq:CostFunction} can be minimized by using different inverse problem strategies such as a truncated SVD algorithm to avoid the divergence due to the noisiest modes. We can therefore invert the $G$ matrix to create the control matrix $G^\dagger$, which is multiplied by the electric field previously estimated with PW to determine the correct voltages $\bar{a}$ to apply, as follows:
\begin{equation}
\label{eq:pasactionneurs}
\bar{a}=-g[\Re (G)^\frown\Im (G)]^{\dagger}[\Re (E_{S}+E_{D})^\frown\Im (E_{S}+E_{D})] ,
\end{equation}
where $^\frown$ means concatenate and $g$ is the servo loop gain, which ensures the loop convergence. In the rest of this paper, $g$ is set to 0.5 to make the correction process as fast as possible while keeping the correction stable with respect to noise.

        \subsection{Simulated on-sky performance}
        \label{subsec:Expectations}
        
\begin{figure}
\centering
\begin{tikzpicture}
        \node (tiger) [anchor=south west, inner sep=0pt] {\includegraphics[width=8.3cm]{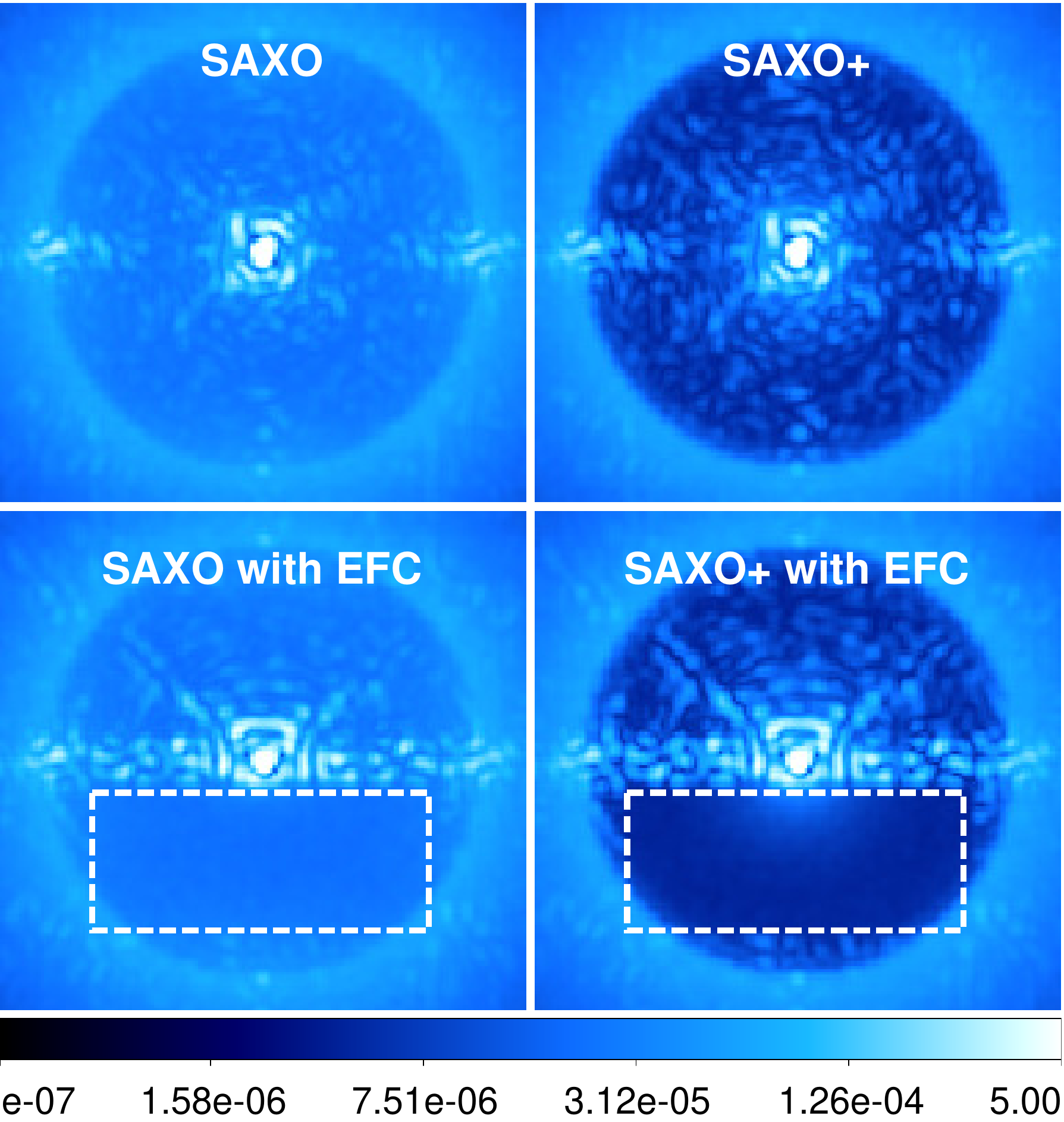}};
        \begin{scope}[x={(tiger.south east)},y={(tiger.north west)}]
        \fill[white] (7.4cm,0cm) rectangle (8.4cm,0.40cm);
    \end{scope}
    \begin{scope}[x={(tiger.south west)},y={(tiger.north east)}]
        \fill[white] (0cm,0cm) rectangle (1.0cm,0.40cm);
    \end{scope}
      \end{tikzpicture}
   \caption[example] 
   { \label{fig:imagesSAXO1SAXO2} 
   Numerical simulations: final coronagraphic images accounting for AO residuals from the current AO system, SAXO (left) and for AO residuals with a faster AO system, SAXO+ (right). Images on the top row are obtained with the APLC alone, while images on the bottom row are obtained after minimization of the stellar speckle intensity using PW+EFC. The HDH region is delimited by the white dashed rectangular box.
   }
\end{figure}

We demonstrate in this section that the SPHERE raw contrast can be significantly improved using our DH strategy via PW and EFC techniques under simulated on-sky conditions. We simulated the expected coronagraphic images obtained with SPHERE under turbulent conditions when applying a PW+EFC solution optimized for the correction $E_D$ and $E_S$ in the HDH region, as described in Sect.~\ref{subsec:PW}.


The target is assumed to be a M0-type star with magnitudes m$_V$=8 and m$_J$=6, observed with the H3 filter ($\lambda_0=1.667\mu$m and $\Delta \lambda=54nm$). The seeing is set to a value of 0.85\,arcsec, with a wind speed of 3\,m/s in three simulated atmospheric layers (located at 0, 1, and 10\,km from the ground). The residual phase aberrations after correction by the AO system are simulated via a CAOS-based numerical tool \citep{Carbillet2008}. This tool uses a power spectral density (PSD) model, including fitting, servo-lag, and aliasing errors. A thousand independent phase screens are randomly drawn from the modeled PSD. A long exposure image is obtained from averaging the frames. Considering the wind speed, the simulated phase screen seen by the VLT pupil becomes incoherent after 2.7\,s. This sequence of a thousand images would therefore correspond to a maximum exposure time of 45 minutes. Two AO systems are simulated: the current SAXO system, based on a SH visible wavefront sensor with 40 sub-pupils across the pupil diameter, and an upgraded system referred to as SAXO+, based on a pyramid IR wavefront sensor to remove the errors due to aliasing and differential refraction \citep{Boccaletti2020}. In the SAXO case, spatial filtering of the SH WFS is taken into account by reducing the aliasing error coefficient by 50\%. Both systems are running at 1.3kHz.

The NCPAs are simulated using the phase screen measured by ZELDA and accounting for 70\,nm rms. The amplitude aberrations are estimated to 8\% rms by taking a pupil image with SPHERE on the calibration unit \citep{Vigan2019}.

The PW+EFC solution is computed to minimize the speckles in the HDH region. The probes used for PW are described in Sect.~\ref{subsec:PW} and are injected with a maximum amplitude of 400\,nm. The EFC optimization computes the correction based on 700 modes. Nine iterations of the PW+EFC process were computed in the absence of turbulence. The final coronagraphic images are simulated including the AO residual turbulent aberrations, the NCPAs and amplitude aberrations, and the PW+EFC optimized HODM phase. All images are normalized by the maximum of the non-coronagraphic point spread function (PSF).

\begin{figure}
\centering
   \includegraphics[width=9cm]{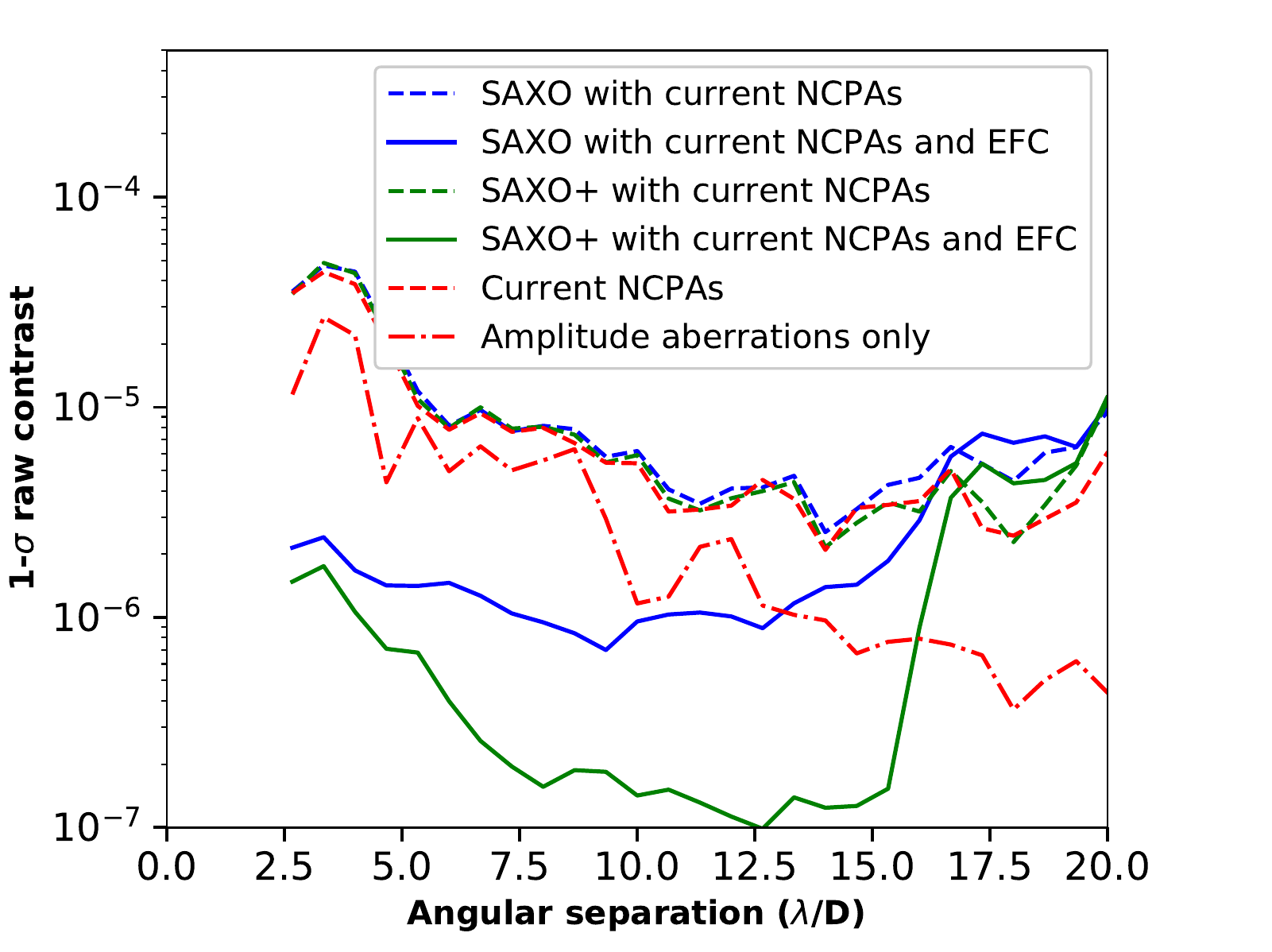}
   \caption[example]
   { \label{fig:SAXO1SAXO2}
   Numerical simulations: 1$\sigma$ contrast as a function of the angular separation. The contrast is  computed in the HDH region in each simulated configuration: with the current SAXO residuals or an upgraded SAXO+ residuals (dashed lines), with and without the EFC phase optimization (solid lines).}
\end{figure}



The final simulated images are shown in Fig.~\ref{fig:imagesSAXO1SAXO2}. The result is optimistic because the numerical simulation assumes an ideal averaging of the turbulent speckles while the NCPAs are perfectly static. The images are therefore NCPA dominated. The PW+EFC correction is applied in a HDH region of size $13\lambda/D\times31\lambda/D$, starting at 2.85$\lambda/D$ away from the central star. Improvement of the contrast appears clearly in the HDH for both simulated AO systems as stellar speckles are no longer detected in the images at the bottom row of the figure.\\
To quantitatively compare the images, the contrast level is computed. We define the contrast as the azimuthal standard deviation limited to the HDH region. The 1$\sigma$ rms contrast curves obtained in the different cases are shown in Fig.~\ref{fig:SAXO1SAXO2}.
Even though SAXO+ provides a better correction below 20\,$\lambda/D$ (the AO cutoff of the 41x41 HODM), our simulations show that the gain with respect to SAXO in contrast is not significant when the NCPAs are not compensated.

We also simulated a perfect cancellation of all phase aberrations. This raw RMS contrast curve sets a contrast limit above $10^{-6}$ at angular distance shorter than 16\,$\lambda/D$ when speckles originating from the amplitude aberrations and the coronagraph residual diffraction pattern are not corrected.

In the case of SAXO, the PW+EFC phase solution leads to a contrast level below the limits set by the NCPAs (red dashed line) or by the amplitude aberrations. The contrast is still limited to a contrast level of $10^{-6}$ by the AO halo that leaves residual speckles in the HDH. For SAXO+, the residual speckles inside the HDH are fainter because the AO system works in a bandwidth close to the science wavelength and is not limited by the NCPAs anymore. The contrast is thus improved especially between 5\,$\lambda/D$ and 16\,$\lambda/D$. As conceived in \cite{Boccaletti2020} for the case of SPHERE+, SAXO+ should also run faster than the current system (2.5kHz) with brighter stars or by using visible photons for low order modes. This would lead to even better contrast performance, especially at low spatial frequencies. However, a complete study of such a system is beyond the scope of this paper.

Compared to the current SAXO performance with no NCPA compensation, applying PW+EFC on SAXO+ should therefore provide a gain of at least one order of magnitude in contrast rms in the HDH region between 3\,$\lambda/D$ and 16\,$\lambda/D$. This improvement can only be reached when applying a strategy that estimates the electric field in the detector plane like PW and corrects in a DH, not only for the NCPAs but for the amplitude aberrations and diffraction residuals from the coronagraph (in this case, EFC). 
\begin{figure*}
\centering
\begin{tikzpicture}
        \node (tiger) [anchor=south west, inner sep=0pt] {\includegraphics[width=18cm]{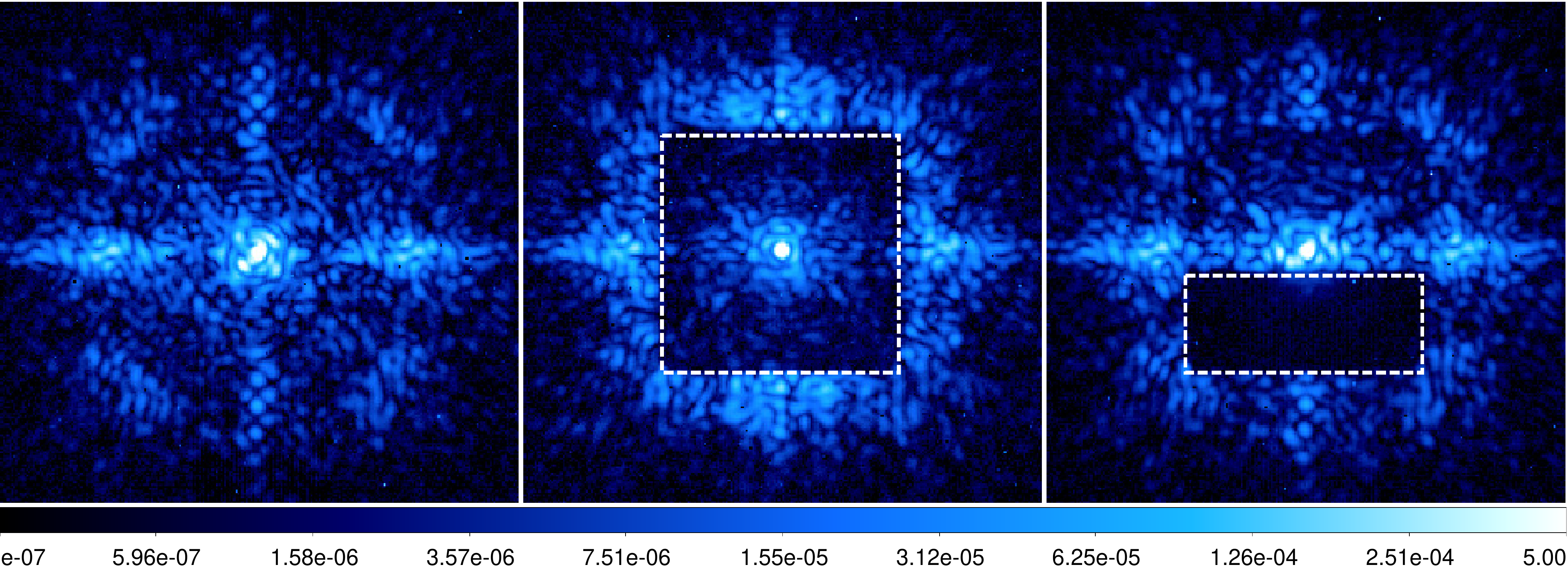}};
        \begin{scope}[x={(tiger.south east)},y={(tiger.north west)}]
        \fill[white] (17cm,0cm) rectangle (18cm,0.4cm);
    \end{scope}
    \begin{scope}[x={(tiger.south west)},y={(tiger.north east)}]
        \fill[white] (0cm,0cm) rectangle (1cm,0.4cm);
    \end{scope}
      \end{tikzpicture}
   \caption[example] 
   { \label{fig:FDHAndHDH_Image} 
   Experimental data: coronagraphic images recorded with SPHERE calibration unit in the H3 band before any NCPA compensation, after 2 iterations of PW+EFC in a FDH (center), and after 6 iterations of PW+EFC in a HDH (right). The sizes of the FDH and HDH are, respectively, $31\lambda/D\times31\lambda/D$ and $13\lambda/D\times31\lambda/D$, 1$\lambda/D$ being equal to $\sim$43~mas. The HDH starts at $2.85\lambda/D$ from the optical axis to get as close as possible from the FPM bottom edge whose radius is 92.5 mas or about $2.15\lambda/D$. The aim is to ensure the stability in the PW+EFC closed loop. The edges of each DH are represented in white dashed lines. The images were normalized by the maximum of the off-axis PSF, recorded at the first iteration of the PW+EFC process.}
\end{figure*}
Thus, any improvement of the SPHERE extreme AO system should include a compensation of amplitude aberrations and a minimization of the coronagraph residuals. Both of these improvements can be achieved with PW+EFC by reducing the DH region to one-half of the image, as demonstrated. Compensation of the amplitude aberrations over the full DH can also be reached by adding another HODM located out of a pupil plane to correct for the amplitude aberrations based on the Talbot effect \citep{Pueyo2009, Beaulieu2017, Baudoz2018SPIE}. Improved optimization of the coronagraph can also minimize the light diffracted by the coronagraph \citep{NDiaye2016feb}.

\section{Wavefront control on the SPHERE internal pupil}
\label{sec:WCinternalpupil}
        \subsection{Strategy}
        \label{subsec:Strategy}

As a proof of concept, we implemented first the PW+EFC technique on SPHERE using the internal source. The associated pupil is different than the VLT pupil, but instead is a simple circular aperture with no central obscuration nor spider supports. The tests were performed with the H3 filter ($\lambda=1667$nm, $\Delta\lambda=54$nm) and 0.83\,s exposures. The PW different probes and the EFC computed correction are not directly applied as voltages on the HODM. Indeed, the SH wavefront sensor would measure an aberration that would be automatically flatten when working in AO closed loop. Instead, we modified the HODM shape by changing the reference slopes of the SH through a daily recorded interaction matrix, which transforms a 1377 voltage vector to a 2480 reference slopes vector \citep[Appendix]{Vigan2019}.


The PW+EFC software we developed is available as an automated package to the support astronomer. The first initialization steps of the procedure include the acquisition of a background image, an off-axis PSF, and a coronagraphic image. The off-axis PSF is dimmed by a neutral density (ND) to avoid saturation. This image is used to normalize the coronagraphic images before multiplication by the model matrix $M^\dagger$ (Eq.~\ref{eq:estimation2}). The first coronagraphic image is taken with a cosine pattern applied on the HODM, creating two bright speckles on both sides of the image center. These spots are used to estimate the center of the H3 images on the IRDIS detector at a subpixel accuracy. This step is crucial to align model images to actual IRDIS images.


The subsequent steps are then repeated iteratively to compute the optimal correction: 1) acquisition of four coronagraphic images using PW probes; 2) computation of the matrix $D$ of Eq.~\ref{eq:DifferenceMatrix} for each pixel $(x,y)$ in the chosen DH region; 3) estimation of the focal plane electric field with Eq.~\ref{eq:estimation2}; 4) multiplication of the estimated field by $G^\dagger$, resulting in the array containing the updated reference slopes for the SH WFS; and 5) acquisition of the corrected coronagraphic image. Each iteration lasts less than two minutes on the internal source.
        
        \subsection{Dark hole creation}
        \label{subsec:DH}
        
The algorithm was applied for different DH configurations. Here, we present two corrections: one in a FDH and one in a HDH.

                \subsubsection{Full dark hole}
                \label{subsubsec:FDH}

\begin{figure}
\centering
   \includegraphics[width=9cm]{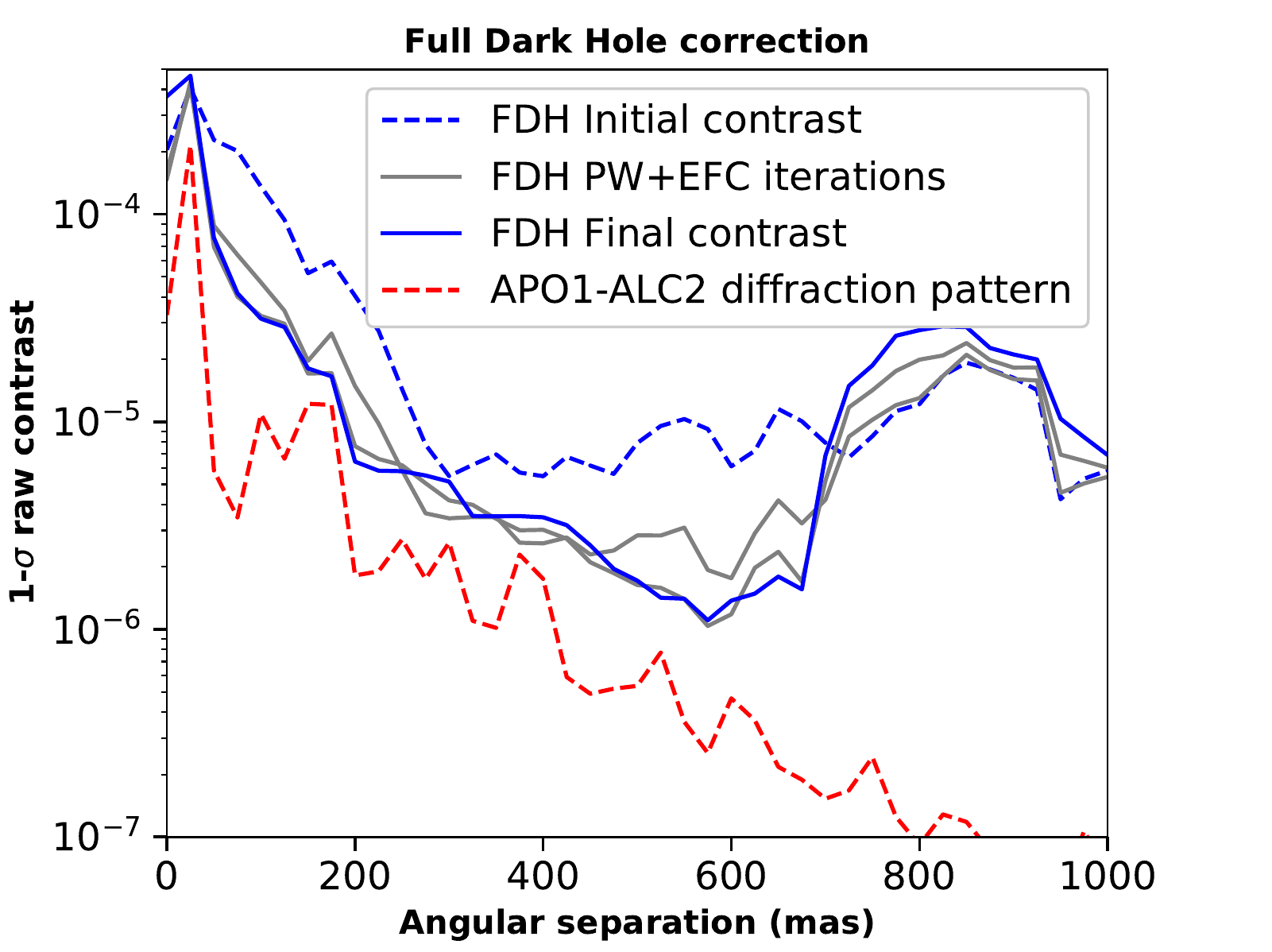}
   \caption[example] 
   { \label{fig:FDH_Contrast} 
   Experimental data: contrast rms during the different iterations of PW+EFC implemented with the SPHERE calibration unit in a FDH of size $31\lambda/D\times31\lambda/D$ (in gray). The curve before compensation is plotted with a blue dashed line; the contrast after the last iteration is represented by a blue continuous line. The theoretical limit set by the APLC diffraction pattern is plotted with a red dashed line.}
\end{figure}

The coronagraphic image that follows the FDH correction is shown at the center of Fig.~\ref{fig:FDHAndHDH_Image}. For this particular experiment, 800 modes of the HODM were used after truncating the SVD in Eq.~\ref{eq:pasactionneurs}. The image can be compared with the image at the left in Fig.~\ref{fig:FDHAndHDH_Image}, which is the initial coronagraphic image before PW+EFC. In order to quantify the improvement, we plot in Fig.~\ref{fig:FDH_Contrast} the \mbox{1$\sigma$} contrast curve before PW+EFC, at each PW+EFC iterations, and after the last iteration. The correction converges after three iterations. The FDH correction improved the contrast in all the spatial frequencies accessible with the HODM by a factor of two to ten, which is close to the results obtained by ZELDA in \cite{Vigan2019}. However, bright speckles are still present near the optical axis in Fig.~\ref{fig:FDHAndHDH_Image}. There are several explanations for this result. First the algorithm is not efficient enough to correct for all the phase aberrations in the DH, especially in a central horizontal bar, as described in Sect.~\ref{subsec:PW}. Second, as explained in Sect.~\ref{subsec:IRDISElectricField}, the image quality is also limited by amplitude aberrations and by the diffraction pattern created by the APLC. We also numerically simulated the raw contrast obtained with a perfect APLC, which is not affected by any aberration (see in Fig.~\ref{fig:FDH_Contrast}, in red dashed lines). We conclude that if the SPHERE APLC has no defects, the performance of PW+EFC in FDH is limited by amplitude aberrations.
                
                \subsubsection{Half dark hole}
                \label{subsubsec:HDH}
\begin{figure}
\centering
   \includegraphics[width=9cm]{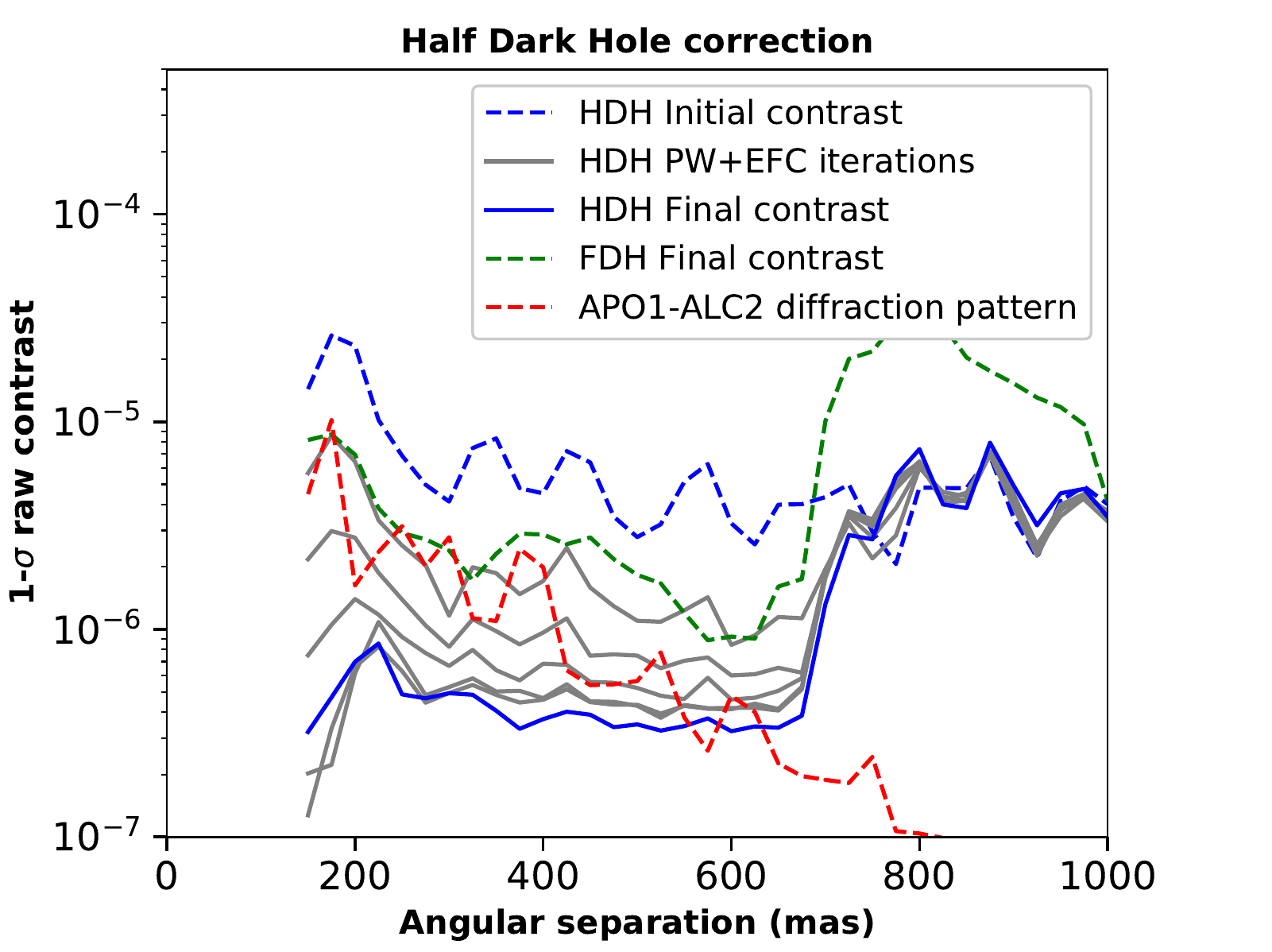}
   \caption[example] 
   { \label{fig:HDH_Contrast} 
   Experimental data: contrast rms during the different iterations of PW+EFC implemented with the SPHERE calibration unit in a HDH of size $13\lambda/D\times31\lambda/D$. In green dashed line is plotted the contrast after the FDH correction (in Sect.~\ref{subsubsec:FDH}) calculated in the HDH region.}
\end{figure}

An alternative way to reach deeper contrast is to sacrifice half the field of view and therefore correct in a HDH region (see in Sect.~\ref{subsec:EFC}). We started the correction from the initial configuration of the HODM as for the FDH (image on the left in Fig.~\ref{fig:FDHAndHDH_Image}). In the right image in Fig.~\ref{fig:FDHAndHDH_Image}, we present the result after six iterations of PW+EFC optimization in a HDH when using 700 correction modes: the correction minimizes all the speckles in the HDH down to the detector noise.
Moreover, we note that the symmetrical region on the other side of the central star (at the top of the image) also benefits from a partial correction due to the compensation of the phase alone. For such a correction, we were required to add a phase pattern of \textasciitilde19nm RMS on the HODM.
The rms contrast curves calculated in the HDH region are plotted in Fig.~\ref{fig:HDH_Contrast}. The deepest contrast is reached after six iterations. At iterations four and five, the contrast is presumably limited by the detector noise (nearly flat contrast floor). This can be overcome with longer integration for each step. Multiplying the exposure time by eight leads to a modest gain in contrast at the sixth iteration. There is room for improvement as long as the wavefront remains stable during acquisition. The HDH reaches a contrast that is significantly better than the theoretical raw contrast of the coronagraph, demonstrating the capability of PW+EFC to go below this limit. At the final step, we measured a contrast below $10^{-6}$ between 150 and 650\,mas, which to our knowledge is the best contrast ever generated on the internal source of SPHERE. 
For demonstration purposes, a HDH was successfully created in the upper part of the image, reaching the same contrast level as that at the bottom of the image shown in Fig.~\ref{fig:FDHAndHDH_Image}. Preliminary experiments also showed the HDH correction was degraded by a factor 2 after two days. Further studies about DH stability are planned.

These results demonstrate that the amplitude aberrations and coronagraph residuals do matter at the $10^{-6}$ level and are critical for future HCI instruments. A compromise would be to sacrifice half of the field of view when using a single DM. With two DMs in the same system, we can consider the FDH correction for instance with the second DM out of the pupil plane, taking advantage of the Talbot effect. An alternative with a single DM would be to generate successively two HDH symmetrically located from the star to cover the entire field of view in two different observing sequences. This solution would need twice as long as the FDH correction with two DM.

\section{Conclusions}
In this paper, we described the contrast limitation due to \mbox{NCPA} and diffraction of the SPHERE APLC coronagraph, which can be overcome with a minor upgrade of the SAXO software and correction strategy. We introduced the PW algorithm, which temporally modulates the speckle intensity in the coronagraphic image to estimate the electric field in the science detector. The PW algorithm only requires four coronagraphic images per iteration. Thanks to numerical simulations, we demonstrated that the PW, in conjunction with EFC control algorithm, is an efficient technique to create deep DH regions under turbulent conditions, by minimizing the speckle intensity due to phase and amplitude aberrations and by reducing the diffraction pattern. We also demonstrated the interest of upgrading the SPHERE AO system to improve the contrast level by a factor of about 10 with respect to what we can reach with the current AO system.

We also reported on experimental results obtained with the calibration unit of SPHERE. Our algorithm PW+EFC used to minimize the star intensity inside a full DH reaches the level set by amplitude aberrations and the diffraction pattern. The same algorithm used to minimize the star intensity in a HDH drastically improves the performance of SPHERE, reaching a contrast level below $10^{-6}$ between 150 mas and 650 mas from the optical axis.


The next step consists in the implementation of the PW+EFC technique on sky during an observation. Two strategies will be tested. First, the NCPA compensation is computed on the SPHERE calibration unit and the resulting correction directly applied on sky. We expect that the performance will not be optimal in that case because the telescope pupil and the internal source pupil differ significantly as well as the amplitude aberrations. The second option consists in computing the correction directly on the targeted star. In that case, long exposures will be needed to average out the atmospheric turbulence phase fluctuations, as explained in \cite{Singh2019} and \cite{PotierAO4ELT2019}, and allow the sensing of the quasi-static aberrations. We also envision experimental tests with other coronagraphs available on SPHERE such as a four-quadrant-phase-mask \citep{Rouan2000} to investigate the potential of the method with smaller IWA coronagraphs.


\begin{acknowledgements}
AP thank the Centre National d'Études Spatiales (CNES) and the DIM-ACAV+ program from the Île-de-France region for funding his PhD work. AV aknowledges funding from the European Research Council (ERC) under the European Union's Horizon 2020 research and innovation programme (grant agreement number 757561). The authors join to thank the reviewer for the constructive remarks provided.
\end{acknowledgements}

\bibliographystyle{aa} 
\bibliography{bib_GS}

\end{document}